\documentclass[conference,10pt]{IEEEtran}
\usepackage{amsmath,amsfonts}
\usepackage{algorithmic}
\usepackage{array}
\usepackage{textcomp}
\usepackage{stfloats}
\usepackage{url}
\usepackage{verbatim}
\usepackage{caption}
\usepackage[labelformat = simple, 
            justification=centering]{subfig}
\usepackage{graphicx}
\usepackage{bm}
\usepackage{cite}
\usepackage{dsfont}
\usepackage{xcolor,soul}
\def\BibTeX{{\rm B\kern-.05em{\sc i\kern-.025em b}\kern-.08em
    T\kern-.1667em\lower.7ex\hbox{E}\kern-.125emX}}
\usepackage{balance}
\begin{document}
\title{A Hybrid Wireless Image Transmission Scheme with Diffusion}

\author{%
  \IEEEauthorblockN{Xueyan Niu\IEEEauthorrefmark{1},
                    Xu Wang\IEEEauthorrefmark{2}\IEEEauthorrefmark{1},
                    Deniz Gündüz\IEEEauthorrefmark{3}\IEEEauthorrefmark{1},
                    Bo Bai\IEEEauthorrefmark{1},
                    Weichao Chen\IEEEauthorrefmark{4},
                    and Guohua Zhou\IEEEauthorrefmark{4}}
  \IEEEauthorblockA{\IEEEauthorrefmark{1}%
                    Theory Lab, Central Research Institute, 2012 Labs, Huawei Technologies Co. Ltd.,
                    \{niuxueyan3, baibo8\}@huawei.com}
  \IEEEauthorblockA{\IEEEauthorrefmark{2}%
                    Department of Computer Science, City University of Hong Kong, Hong Kong SAR, China,
                    xu.wang@my.cityu.edu.hk}
  \IEEEauthorblockA{\IEEEauthorrefmark{3}%
                    Department of Electrical and Electronic Engineering, Imperial College London, London, UK,
                    d.gunduz@imperial.ac.uk}
  \IEEEauthorblockA{\IEEEauthorrefmark{4}%
                    RAN Research Department WN, Huawei Technologies Co. Ltd., 
                    carlyle.chen@tongji.edu.cn, guohua.zhou@huawei.com}
}

\markboth{Journal of \LaTeX\ Class Files,~Vol.~18, No.~9, September~2020}%
{How to Use the IEEEtran \LaTeX \ Templates}

\maketitle

\begin{abstract}
We propose a hybrid joint source-channel coding (JSCC) scheme, in which the conventional digital communication scheme is complemented with a generative refinement component to improve the perceptual quality of the reconstruction. The input image is decomposed into two components: the first is a coarse compressed version, and is transmitted following the conventional separation based approach. An additional component is obtained through the diffusion process by adding independent Gaussian noise to the input image, and is transmitted using DeepJSCC. The decoder combines the two signals to produce a high quality reconstruction of the source. Experimental results show that the hybrid design provides bandwidth savings and enables graceful performance improvement as the channel quality improves.
\end{abstract}


\section{Introduction}

The fast increasing demand for wireless transmission of high-resolution image and video signals poses a challenge to current communication systems, as emerging applications such as metaverse, augmented/virtual reality (AR/VR), Internet-of-things (IoT), vehicular-to-everything (V2X), require more robust transmission and realistic reconstruction of video in a fast-varying wireless communication environment with limited bandwidth resources. 
State-of-the-art (SOTA) digital communication systems are designed based on Shannon's source-channel separation theorem \cite{Shannon}, which implies that there is no loss of optimality by applying separate source coding followed by channel coding, in the asymptotic infinite block length regime and for ergodic source and channel statistics. In reality, these idealized assumptions are rarely met \cite{Goldsmith1995}; and therefore, the separation-based digital communication systems do not operate at the theoretical optimal \cite{verdu1995}, especially in the finite block-length regime \cite{Kostina2013}. Moreover, separation-based digital communication suffers from sudden quality drop when the channel (signal-to-noise ratio) SNR drops below a certain threshold, known as the ``cliff effect'', which requires operating well below the instantaneous channel capacity over time-varying wireless channels.

Joint source-channel coding (JSCC) has long been studied as an alternative approach to improve the end-to-end performance in practical systems. Indeed, JSCC predates separation based digital transmission approaches, as analog and frequency modulation (AM/FM) are JSCC schemes based on direct modulation of the continuous-time input signal onto the carrier waveform. Later, also in the discrete-time communication framework, JSCC has been shown to outperform purely separate approaches in image and video transmission tasks, particularly in the limited bandwidth scenarios and to provide more resilience to channel variations \cite{Goldsmith1995, Zhai2005}. More recently, in the context of semantic communications \cite{Beyond}, deep learning based JSCC methods, e.g., DeepJSCC, have shown remarkable results thanks to their ability to learn the mapping directly from the training data (for both source and channel) \cite{DeepJSCC, kurka:TWC:21, tung2021deepwive, Wang:SPL:21, yang2022ofdm, Shao:WCL:23, Wu:WCL:22}. Unlike the separation-based digital transmission schemes, JSCC-based methods directly map the image pixel values to channel input symbols. Through end-to-end training, the encoder and decoder pair learn to operate under various channel conditions. 


The hybrid communication scheme proposed in this paper envisions a system that inherits the advantages of both digital and joint encoding schemes. By integrating the JSCC-based communication into the digital communication infrastructure, which has already been widely deployed, this method aims to provide bandwidth savings while delivering content with higher perceptual quality more robustly over unreliable wireless channels. We send a low-resolution digitally compressed version of the input image first by following the conventional separation-based digital communication approach. Then, we send a refinement component obtained through the diffusion process using DeepJSCC \cite{DeepJSCC}, to improve the perceptual quality of the reconstructed image. Inspired by the success of a class of image generation techniques known as diffusion models \cite{pmlr-v37-sohl-dickstein15,NEURIPS2020_4c5bcfec}, in particular, the score-based diffusion models \cite{NEURIPS2019_3001ef25,NEURIPS2020_92c3b916}, the refinement information is obtained by slowly adding white noise to the signal such that the source distribution is transformed to a Gaussian shape after the Markov chain of diffusion steps. Compared to other image generation methods, notably generative adversarial networks (GANs), diffusion based image generation exhibits better image sample quality \cite{NEURIPS2021_49ad23d1}. 
Moreover, since the diffusion process results in an approximately Gaussian signal, we exploit the optimality of `analog/uncoded' transmission of Gaussian sources over Gaussian channel \cite{Goblick:TIT:65}, and transmit this part using JSCC. Experimental results show that using the same bandwidth and power resources, compared to using only digital transmission, the proposed method achieves performance gain in terms of the reconstruction quality while also providing graceful improvement of the performance as the channel SNR increases, while the quality of the pure digital transmission does not increase once the compression rate is fixed.

\begin{figure*}[t]
\centering
\includegraphics[width=6.5in]{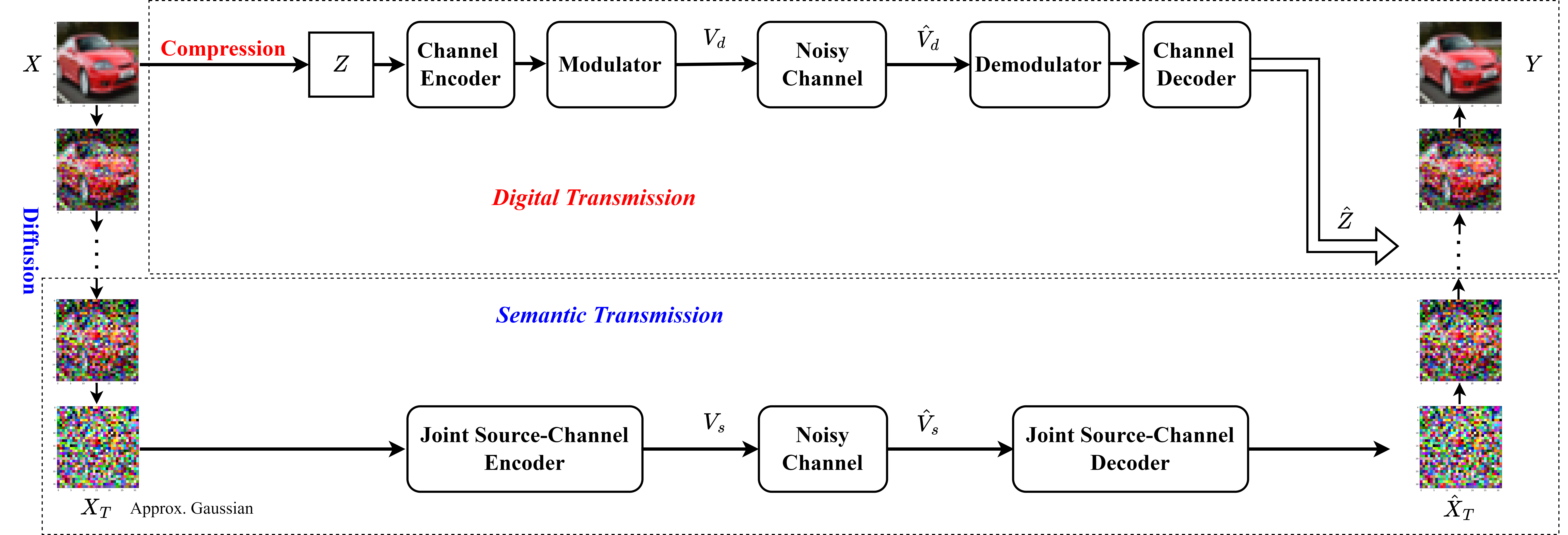}
\caption{Illustration of the proposed hybrid image transmission scheme. The upper part shows the digital transmission component, where a coarse compressed version $Z$ of the input image $X$ is digitally transmitted over the wireless channel. By applying diffusion steps, the noisy version of the source signal $X_T$ is extracted, which approximately follows a Gaussian distribution. The lower part in the figure shows JSCC of $X_T$ over the channel. The two signals $\hat{Z}$ and $\hat{X}_T$ are then combined to generate reconstruction $Y$ using the reverse diffusion steps. The digital stream ensures a reasonable accuracy under distortion metrics, while the refinement stream aims to improve the perceived visual quality.}
\label{fig:framework}
\end{figure*}


\section{Problem Formulation}

We consider image transmission over a wireless channel with limited bandwidth and a transmitter power constraint. Consider images of height $H$, width $W$, and $C$ color channels. The input image is represented by a real-valued vector $\bm x\in \mathbb{R}^n,$ where $n=H\cdot W\cdot C.$ The transmitter maps the input image $\bm x$ into a complex-valued vector $\bm v\in \mathbb{C}^k$ to be transmitted over the noisy channel. The ratio $\rho= k/n$ is defined as the \textit{bandwidth ratio} in the JSCC literature, which indicates the average number of channel symbols available for each source symbol. 
We use capital letters such as $X$ to denote random variables, lower-case letters such as $\bm x$ to denote corresponding (vector) instances.
In practice, an average power constraint is also imposed on the transmitter: $1/k \mathds{E}[V V^*] \leq 1$. Let $\hat{\bm v}\in \mathbb{C}^k$ denote the channel output corrupted by channel noise. The receiver estimates the input image based on $\hat{\bm v}$. Let $\hat{\bm x}\in \mathbb{R}^n$ denote the reconstructed image at the receiver. The quality of the reconstruction is measured by some specified distortion measure between the original image and the reconstruction. The goal of wireless image transmission is to design a system that optimizes the performance of the reconstruction under limited bandwidth and power resources.

\subsection{Separation-Based Digital Transmission}
In current digital transmission systems, image compression and channel coding are separately performed. The source-encoded data is transmitted through the wireless channel after channel coding and modulation. Images are first compressed using established codecs such as JPEG and JEPG2000, which consist of sequentially applying some transform coding to the image pixels, e.g., discrete cosine transform (DCT) or discrete wavelet transform (DWT), followed by quantization and entropy coding. Channel coding follows immediately, as an ideal source coding is not resilient to channel errors. SOTA channel codes include Turbo, low density parity check (LDPC) and polar codes. These codes are known to perform close to the Shannon capacity in the large blocklength regime. The encoded bitstream is then mapped to some discrete input constellation, such as 16-QAM and 64-QAM, which maps the bit sequence to complex-valued channel symbols to be transmitted over the wireless channel. 

The receiver reverses these procedures by first demodulating and decoding the channel code, trying to mitigate any impact of the channel noise, and the decompressor is applied afterwards to reconstruct the original input image. The demodulator, channel decoder, and decompressor are chosen to match the forward modules in the encoding process. The source and channel coding rates and the modulation scheme are chosen jointly according to the channel condition and the source characteristics to minimize the end-to-end distortion, which is caused by both the errors over the channel and the quantization in source coding.

\section{Hybrid Transmission Framework}
In this section, we will introduce the proposed hybrid transmission scheme that benefits from both the accuracy of the separation-based digital communication system and the efficiency and robustness of the JSCC scheme.

\subsection{Model Description}
A diagram of the system model is shown in Fig. \ref{fig:framework}. Consider the input signal $X$ with sample space consisting of images (real-valued vectors) in $\mathbb{R}^n.$ In the hybrid framework, the signal $X$ is decomposed into the pair $(Z, X_T)=f_{\theta}(X),$ where $Z$ represents a generic compressed version of $X$ such that $H(Z) \ll H(X_0),$ i.e., the number of bits required to represent $Z$ is much smaller than that for $X$. 

The coarse compressed component $\bm z$ is transmitted in the conventional digital manner  (e.g., LDPC code + 16-QAM) to obtain a complex-valued channel input $\bm v_d = f_d(\bm z)\in \mathbb{C}^{k_d},$ where $k_d$ is the dimension of the channel input for digital transmission. 

The complementary component $X_T$ is obtained by following a forward diffusion process, where $X_0= X$, and $X_t$ is corrupted from $t=0$ to $t=T$ using independent additive Gaussian noise at each step, so that $\bm x_T\in \mathbb{R}^{k_s}$ approximately follows a Gaussian distribution. Moreover, by integrating convolutional layers into the neural network of the diffusion model, the dimension of the data is reduced. This final result of the diffusion process is transmitted directly over the channel, by first pairing the real outputs to form complex channel inputs. We denote the corresponding channel input by $\bm v_s \in \mathbb{C}^{k_s}$. Overall, the channel input is obtained by the concatenation of the digital and diffusion-based joint encoded components, $V = [V_d V_s]$, for which the bandwidth ratio is given by $\rho = (k_d + k_s)/n$. We also allocate the available power between the two streams $V_d$ and $V_s$ to optimize the performance. 

Through end-to-end training, the decoder learns a reverse diffusion process that recovers the signal at $t=0$ from $t=T$. So, after receiving $(\hat{V}_d, \hat{V}_s),$ the receiver first recovers $\hat{Z}$ and $\hat{X}_T$, and then generates a reconstruction $Y = g_{\phi}(\hat{Z}, \hat{X}_T),$ where $g_\phi$ is a pre-trained neural network reversing a conditional diffusion process.

\begin{figure*}[!t]
  \centering
  \includegraphics[width=6in]{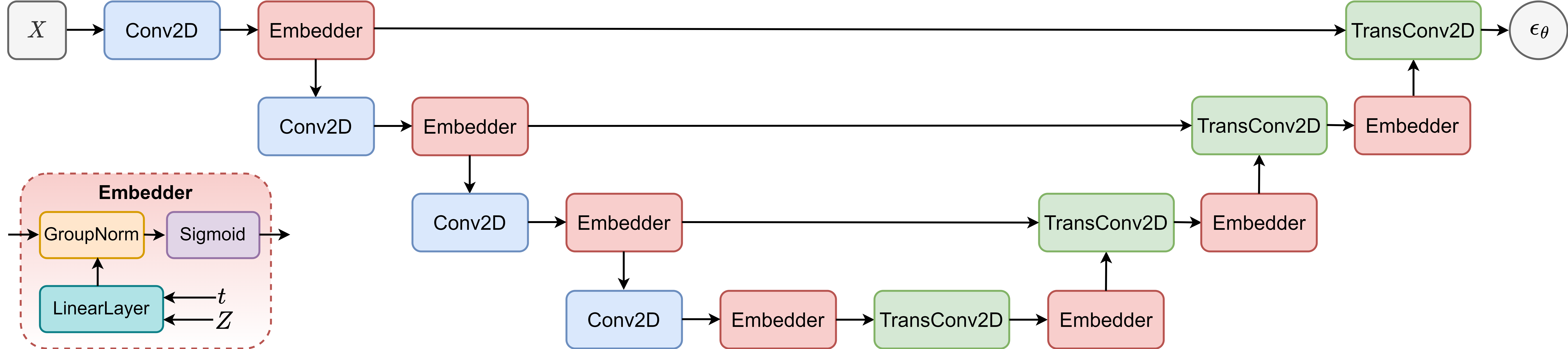}
  \caption{The neural network architecture of the JSCC encoder and decoder.}
  \label{fig:NN}
\end{figure*}

\subsection{Decomposition of the Source Signal}
In principle, the compression $Z$ can be obtained using an arbitrary compression scheme. Arguably, the most common image compression algorithm is the JPEG standard. When applying the JPEG compression, the input image is first divided into small tiles, then the DCT transform is applied, and the resulting coefficients are quantized with a pre-defined quantization table. The level of quantization can be chosen to achieve different reconstruction qualities. When less number of bits are used, the reconstructed image becomes more blurred. In our setting, $Z$ can be a coarse compression of $X$ with very low number of bits per pixel.

Recent research shows that the JSCC scheme combined with a generative model for reconstruction can achieve significant bandwidth reduction, while significantly improving the perceptual quality of the reconstruction \cite{Erdemir22}. While a pretrained generative model based on GANs is employed in \cite{Erdemir22}, here we will use a diffusion process, which has shown remarkable generative capability in a series of recent papers \cite{NEURIPS2019_3001ef25,NEURIPS2020_92c3b916}.  

The forward diffusion process is undertaken to encode the refinement information with the following Gaussian transition kernel:
\begin{align}
    p_t(\bm x_t | \bm x_{t-1}) = \mathcal{N}(\bm x_t; \sqrt{1-\beta_t}\bm x_{t-1}, \beta_t \bm I). 
\end{align}

Furthermore, $X_t$ can be sampled directly according to the cumulative kernal \cite{NEURIPS2020_4c5bcfec}, such that 
\begin{align}
    X_t = \sqrt{\bar{\alpha}_t}X_0 + \sqrt{1-\bar{\alpha}_t}\bm \epsilon, 
\end{align}
where $\bar{\alpha}_t = \prod_{s=1}^t(1-\beta_s)$ and $\bm \epsilon\sim \mathcal{N}(\bm 0, \bm I).$

\subsection{Channel Transmission}
The additive white Gaussian noise (AWGN) channel is adopted in this work, as it has been widely used to represent realistic wireless channel conditions. 
The channel input signals are transmitted through the noisy channel with the following transfer function
\begin{align}
    \eta_n(V) = V + \bm n,
\end{align}
where $\bm n$ is the additive independent and identically distributed (i.i.d.) Gaussian noise signal, $\bm n \sim \mathcal{CN}(\bm 0, \sigma^2\bm I),$ and $\sigma^2$ is the average noise power. We enforce a total average power constraint such that
\begin{align}
    \frac{1}{n} \mathds{E} [{V}_d {V}_d^* + {V}_s {V}_s^*] \leq  1.
\end{align}
The quality of the communication channel is measured by the average SNR, defined as 
$
\mathrm{SNR} = 10 \log_{10}\frac{1}{\sigma^2}.
$


Notably, since the signal $X_T$ approximately follows a Gaussian distribution, it is expected that transmitting it over the AWGN in an `analog/uncoded' fashion is more efficient, since it is known that the uncoded transmission of i.i.d. Gaussian samples over an AWGN channel achieves the optimal performance despite operating over a finite block lengnth \cite{Goblick:TIT:65}. Here, instead of the channel coding/decoding and channel modulation/demodulation, a pair of joint source-channel encoder and decoder is trained in an end-to-end fashion, treating the AWGN channel as a non-trainable layer represented by the transfer function $\eta_n$ with a range of SNR values.


\captionsetup[subfigure]{labelformat=empty}

\begin{figure}[!t]
\centering
\subfloat[PSNR/SSIM]{
\includegraphics[width=2.5cm]{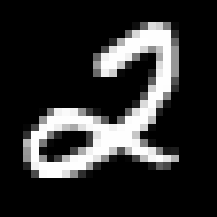}
}
\subfloat[18.77dB/0.76]{
\includegraphics[width=2.5cm]{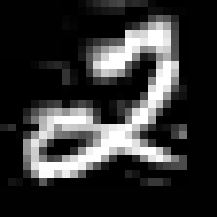}
} 
\subfloat[20.47dB/0.82]{
\includegraphics[width=2.5cm]{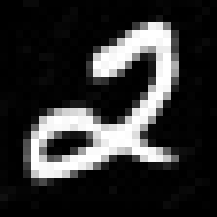}
}\\
\subfloat[PSNR/SSIM]{
\includegraphics[width=2.5cm]{new_figs/original.png}
}
\subfloat[19.91dB/0.77]{
\includegraphics[width=2.5cm]{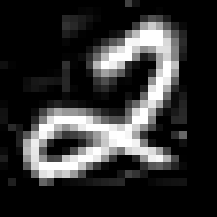}
}
\subfloat[21.85dB/0.84]{
\includegraphics[width=2.5cm]{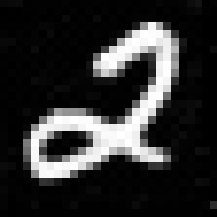}
}\\
\subfloat[PSNR/SSIM\\ (a) Original data $X$]{
\includegraphics[width=2.5cm]{new_figs/original.png}
}
\subfloat[21.13dB/0.81\\ (b) Digital part $\hat{Z}$]{
\includegraphics[width=2.5cm]{new_figs/jpeg6.png}
}
\subfloat[22.85dB/0.86\\ (c) Reconstruction $Y$]{
\includegraphics[width=2.5cm]{new_figs/diff6.png}
}
\caption{Example of reconstructions of (a) original images produced by (b) the baseline digital schemes that concatenate JPEG image compression, LDPC code, and QAM modulation, and (c) our hybrid scheme.
From top to bottom, the rows correspond to bandwidth compression ratios 1/4, 
3/8, 5/8.}
\label{fig:mnist}
\end{figure}



\subsection{Neural Network Architecture}
As shown in Fig.~\ref{fig:NN}, for the diffusion model, we use the common U-net architecture \cite{u-net} with adaptations \cite{NEURIPS2019_3001ef25}, which consists of multiple 2D convolution layers. We use positional embedding to encode the time step $t.$ Each embedder block consists of a goup norm, a sigmoid block, and a linear layer that incorporates $t$ and the conditional information from the digital transmission.

The objective of the training is to obtain a high-quality reconstruction $\bm y\sim P_Y$ of the realization $\bm x_0\in \mathbb{R}^{n}$ in the same sample space at the decoder's end. The quality of the reconstruction is traditionally evaluated using distortion measures such as the peak signal-to-noise ratio (PSNR). Other measures, such as the structural similarity index (SSIM), and the learned perceptual image patch similarity (LPIPS) have been shown to better capture the perceptual quality of the construction, which is a major focus of semantic communications \cite{Beyond}.

In a recently developed theory of the rate-distortion-perception trade-off \cite{pmlr-v97-blau19a,niu2023conditional,hamdi2023rate}, the perceptual quality is measured by the discrepancy between the probability distributions of the input data and the reconstruction. In addition, the semantic information $X_T$ can be viewed as a latent variable, following the line of research in \cite{9844779}. Therefore, the model is trained to minimize the average distortion between the input $X$ and its reconstructions $Y$ as well as the distance between the input distribution $P_{X}$ and the output distribution $P_Y$ capturing the perceptual quality, i.e.,
\begin{align}
\min_{\theta, \phi}  \lambda_1\mathds{E}_{p(\bm x,\bm y)} [d(X, Y)]
+ \lambda_2 \mathcal{L}(P_X, P_Y), 
\end{align}
where $d(\cdot,\cdot)$ and $\mathcal{L}(\cdot,\cdot)$ represent the distortion metric and the perceptual loss. The digital transmission stream ensures a reasonable accuracy of the reconstruction using distortion metrics, while the forward and reverse diffusion processes operates directly on the probability distributions, which improve the perceived visual quality of the reconstruction. 

\begin{figure} [t]
  \centering
\subfloat[(a) PSNR]{%
     \includegraphics[height=4.2cm]{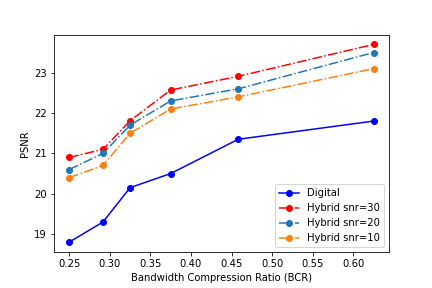}}
  \hfill
\subfloat[(b) SSIM]{%
      \includegraphics[height=4.2cm]{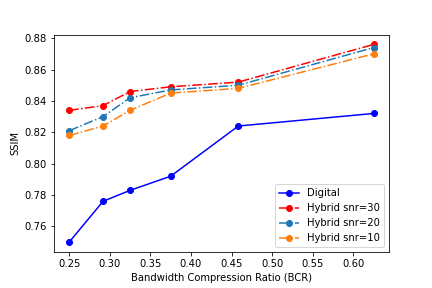}}
  \hfill
\subfloat[(c) LPIPS]{%
      \includegraphics[height=4.2cm]{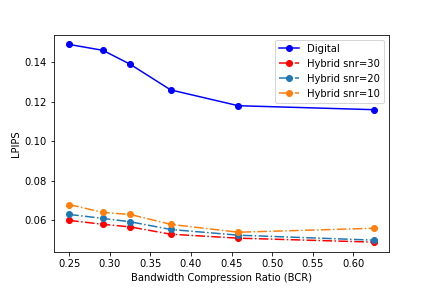}}
\caption{Comparison of the proposed hybrid transmission scheme trained under channel $\mathrm{SNR}=10 \mathrm{dB}$ with the baseline digital scheme evaluated using (a) PSNR (larger is better), (b) SSIM (larger is better), and (c) LPIPS (smaller is better) over various channel conditions with $\mathrm{SNR}=10,20,30 (\mathrm{dB}).$}
\label{fig:results} 
\end{figure}

\section{Simulation Results}
We evaluate the performance of the proposed hybrid digital-semantic communication framework under difference channel SNRs on the MNIST database, which contains 60,000 training images and 10,000 testing images of hand-written digits. The dimension of the images is $N=28 \times 28 \times 1$ (height, width, channels). The first column of Fig.~\ref{fig:mnist} are examples of the original images. 


For the digital transmission stream, we concatenate the JPEG compression with LDPC codes, QAM modulation, and AWGN channel sequentially.
We implemented combinations of 4-QAM, 16-QAM,
and 64-QAM modulation schemes and LDPC codes with corresponding rates.
Examples of recovered images after compression, channel coding, modulation, and their reversals are shown in the second column of Fig.~\ref{fig:mnist}. For fair comparison, we stripped the header information for JPEG when computing the source coding rates.

For the semantic transmission stream, we train the model on the AWGN channel with $\mathrm{SNR}=10 \mathrm{dB}.$ The batchsize during the training is set as 64 and the learning rate is $1e^{-4}$. 
The reconstructed images at the receiver combining the digital and semantic datastreams are shown in the third column of Fig.~\ref{fig:mnist}.

We further test the pre-trained system under different channel conditions with $\mathrm{SNR}=10, 20, 30 (\mathrm{dB}).$ The results are presented in Fig.~\ref{fig:results}.
At the same bandwidth compression level, the hybrid scheme significantly improves the reconstruction quality in terms of PSNR, SSIM, and LPIPS. 
In comparison to the digital transmission scheme, when $\mathrm{PSNR}=21,$ the hybrid scheme provides a bandwidth reduction of 33.3\%; when $\mathrm{SSIM}=0.83,$ the hybrid scheme provides a bandwidth reduction of 47.2\%.
Moreover, when testing under different channel SNRs (dashed lines), the performances do not suffer from the ``cliff effect'', which indicates an improved robustness of the transmission under channel variation.

We note here that the current results are limited to the MNIST dataset mainly due to the difficulty of training the neural network associated with the diffusion model. These should be treated as promising initial results, and more complex datasets using more efficient training techniques is currently under investigation.


\section{Conclusion and Future Work}
We propose a novel image transmission scheme that combines the SOTA digital communication with the emerging semantic communication utilizing recent developments in diffusion-based generative modeling. The hybrid scheme provides bandwidth savings while providing graceful performance improvement with channel SNR. There are several interesting directions for future research. First, current hybrid framework is designed for and evaluated on AWGN channels. Future investigations will include extensions to other channel models, including fading channels. Second, the proposed algorithm is designed for image transmission. In principle, other types of data, such as video and audio, can also be transmitted using the same framework. Third, efficient algorithms for power allocation between the digital and semantic signals with power division rather than time division shall be investigated.

\bibliographystyle{IEEEtran}
\bibliography{ref}

\begin{thebibliography}{10}
\providecommand{\url}[1]{#1}
\csname url@samestyle\endcsname
\providecommand{\newblock}{\relax}
\providecommand{\bibinfo}[2]{#2}
\providecommand{\BIBentrySTDinterwordspacing}{\spaceskip=0pt\relax}
\providecommand{\BIBentryALTinterwordstretchfactor}{4}
\providecommand{\BIBentryALTinterwordspacing}{\spaceskip=\fontdimen2\font plus
\BIBentryALTinterwordstretchfactor\fontdimen3\font minus
  \fontdimen4\font\relax}
\providecommand{\BIBforeignlanguage}[2]{{%
\expandafter\ifx\csname l@#1\endcsname\relax
\typeout{** WARNING: IEEEtran.bst: No hyphenation pattern has been}%
\typeout{** loaded for the language `#1'. Using the pattern for}%
\typeout{** the default language instead.}%
\else
\language=\csname l@#1\endcsname
\fi
#2}}
\providecommand{\BIBdecl}{\relax}
\BIBdecl

\bibitem{Shannon}
C.~E. Shannon, ``A mathematical theory of communication,'' \emph{The Bell
  System Technical Journal}, vol.~27, no.~3, pp. 379--423, 1948.

\bibitem{Goldsmith1995}
A.~Goldsmith, ``Joint source/channel coding for wireless channels,'' in
  \emph{IEEE Vehicular Techn. Conf.}, 1995, pp. 614--618.

\bibitem{verdu1995}
S.~Vembu, S.~Verdu, and Y.~Steinberg, ``The source-channel separation theorem
  revisited,'' \emph{IEEE Trans. Inf. Theory}, vol.~41, no.~1, 1995.

\bibitem{Kostina2013}
V.~Kostina and S.~Verdú, ``Lossy joint source-channel coding in the finite
  blocklength regime,'' \emph{IEEE Trans. Inf. Theory}, vol.~59, no.~5, 2013.

\bibitem{Zhai2005}
F.~Zhai, Y.~Eisenberg, and A.~Katsaggelos, \emph{Joint Source-Channel Coding
  for Video Communications}.\hskip 1em plus 0.5em minus 0.4em\relax Elsevier
  Inc, Dec. 2005, pp. 1065--1082.

\bibitem{Beyond}
D.~Gündüz, Z.~Qin, I.~E. Aguerri, H.~S. Dhillon, Z.~Yang, A.~Yener, K.~K.
  Wong, and C.-B. Chae, ``Beyond transmitting bits: Context, semantics, and
  task-oriented communications,'' \emph{IEEE Journal on Selected Areas in
  Communications}, vol.~41, no.~1, pp. 5--41, 2023.

\bibitem{DeepJSCC}
E.~Bourtsoulatze, D.~Burth~Kurka, and D.~Gündüz, ``Deep joint source-channel
  coding for wireless image transmission,'' \emph{IEEE Trans. on Cognitive
  Comms. and Netw.}, vol.~5, no.~3, pp. 567--579, 2019.

\bibitem{kurka:TWC:21}
D.~B. Kurka and D.~G{\"u}nd{\"u}z, ``Bandwidth-agile image transmission with
  deep joint source-channel coding,'' \emph{IEEE Transactions on Wireless
  Communications}, vol.~20, no.~12, pp. 8081--8095, 2021.

\bibitem{tung2021deepwive}
T.-Y. Tung and D.~Gündüz, ``{DeepWiVe}: Deep-learning-aided wireless video
  transmission,'' \emph{IEEE Journal on Selected Areas in Communications},
  vol.~40, no.~9, pp. 2570--2583, 2022.

\bibitem{Wang:SPL:21}
M.~Wang, Z.~Zhang, J.~Li, M.~Ma, and X.~Fan, ``Deep joint source-channel coding
  for multi-task network,'' \emph{IEEE Signal Processing Letters}, vol.~28, pp.
  1973--1977, 2021.

\bibitem{yang2022ofdm}
M.~Yang, C.~Bian, and H.-S. Kim, ``{OFDM}-guided deep joint source channel
  coding for wireless multipath fading channels,'' \emph{IEEE Transactions on
  Cognitive Communications and Networking}, 2022.

\bibitem{Shao:WCL:23}
Y.~Shao and D.~Gunduz, ``Semantic communications with discrete-time analog
  transmission: A {PAPR} perspective,'' \emph{IEEE Wireless Communications
  Letters}, vol.~12, no.~3, pp. 510--514, 2023.

\bibitem{Wu:WCL:22}
H.~Wu, Y.~Shao, K.~Mikolajczyk, and D.~Gündüz, ``Channel-adaptive wireless
  image transmission with {OFDM},'' \emph{IEEE Wireless Communications
  Letters}, vol.~11, no.~11, pp. 2400--2404, 2022.

\bibitem{pmlr-v37-sohl-dickstein15}
J.~Sohl-Dickstein, E.~Weiss, N.~Maheswaranathan, and S.~Ganguli, ``Deep
  unsupervised learning using nonequilibrium thermodynamics,'' in \emph{Int'l
  Conf. on Machine Learning (ICML)}, Jul 2015.

\bibitem{NEURIPS2020_4c5bcfec}
J.~Ho, A.~Jain, and P.~Abbeel, ``Denoising diffusion probabilistic models,'' in
  \emph{Advances in Neural Info. Proc. Sys. (NeurIPS)}, 2020.

\bibitem{NEURIPS2019_3001ef25}
Y.~Song and S.~Ermon, ``Generative modeling by estimating gradients of the data
  distribution,'' in \emph{Adv. in Neural Inf. Proc. Sys. (NeurIPS)}, 2019.

\bibitem{NEURIPS2020_92c3b916}
------, ``Improved techniques for training score-based generative models,'' in
  \emph{Advances in Neural Inf. Proc. Sys. (NeurIPS)}, 2020.

\bibitem{NEURIPS2021_49ad23d1}
P.~Dhariwal and A.~Nichol, ``Diffusion models beat {GAN}s on image synthesis,''
  in \emph{Advances in Neural Inf. Proc. Sys. (NeurIPS)}, 2021, pp. 8780--8794.

\bibitem{Goblick:TIT:65}
T.~Goblick, ``Theoretical limitations on the transmission of data from analog
  sources,'' \emph{IEEE Trans. Inf. Theory}, vol.~11, no.~4, 1965.

\bibitem{Erdemir22}
E.~Erdemir, T.-Y. Tung, P.~L. Dragotti, and D.~Gunduz, ``Generative joint
  source-channel coding for semantic image transmission,'' \emph{arXiv}, 2022.

\bibitem{u-net}
O.~Ronneberger, P.~Fischer, and T.~Brox, ``U-net: Convolutional networks for
  biomedical image segmentation,'' in \emph{Medical Image Comp. and
  Computer-Assisted Inter.}, 2015.

\bibitem{pmlr-v97-blau19a}
Y.~Blau and T.~Michaeli, ``Rethinking lossy compression: The
  rate-distortion-perception tradeoff,'' in \emph{Int'l Conf. on Mach. Learning
  (ICML)}, Jun 2019, pp. 675--685.

\bibitem{niu2023conditional}
X.~Niu, D.~G{\"u}nd{\"u}z, B.~Bai, and W.~Han, ``Conditional
  rate-distortion-perception trade-off,'' in \emph{2023 IEEE International
  Symposium on Information Theory (ISIT)}.\hskip 1em plus 0.5em minus
  0.4em\relax IEEE, 2023, pp. 1074--1079.

\bibitem{hamdi2023rate}
Y.~Hamdi and D.~G{\"u}nd{\"u}z, ``The rate-distortion-perception trade-off with
  side information,'' in \emph{2023 IEEE International Symposium on Information
  Theory (ISIT)}.\hskip 1em plus 0.5em minus 0.4em\relax IEEE, 2023, pp.
  1056--1061.

\bibitem{9844779}
J.~Liu, S.~Shao, W.~Zhang, and H.~V. Poor, ``An indirect rate-distortion
  characterization for semantic sources: General model and the case of gaussian
  observation,'' \emph{IEEE Trans. Comms.}, vol.~70, no.~9, 2022.

\end{thebibliography}

\begin{IEEEbiographynophoto}{Jane Doe}
Biography text here without a photo.
\end{IEEEbiographynophoto}

\begin{IEEEbiography}[{\includegraphics[width=1in,height=1.25in,clip,keepaspectratio]{fig1.png}}]{IEEE Publications Technology Team}
In this paragraph you can place your educational, professional background and research and other interests.\end{IEEEbiography}

\end{document}